\documentclass[aps,twocolumn,superscriptaddress,floatfix,longbibliography,prl]{revtex4-1}
\usepackage{tikz}
\usepackage{lipsum}
\usepackage{graphicx}
\usepackage[export]{adjustbox}
\usepackage{amsmath,amssymb,wasysym}
\usepackage{braket}
\usepackage{mathtools}
\usepackage{mathrsfs}
\usepackage{verbatim}
\usepackage{makecell}
\usepackage{cases}
\usepackage{bbm}
\usepackage{hyperref}

\newcommand{\im}{\mathbf{i}}

\newcommand{\sech}{\mathrm{sech}}
\begin{document}
\title{Comment: ``Chaotic-Integrable Transition in the Sachdev-Ye-Kitaev Model''}

\author{Jaewon Kim}
\author{Xiangyu Cao}
\affiliation{Department of Physics, University of California, Berkeley, CA 94720, USA}
\maketitle

In the Letter~\cite{SYK24} the authors studied a variant of Sachdev-Ye-Kitaev model with both quadratic and quartic interactions (also known as mass-deformed SYK). They claimed that the quantum Lyapunov exponent $\lambda_L$ would vanish below some critical temperature. In this Comment we show that this is not the case. We calculate $\lambda_L$ exactly in the perturbative regime where the temperature $T$ and the quartic coupling $J$ are both much smaller than the quadratic coupling $\kappa$. At leading order in $T/\kappa$ and $ J/\kappa$, we find
\begin{equation}
    \frac{\lambda_L}{\kappa} = \frac{3T^2 J^2}{\kappa^4} \,. \label{eq:main}
\end{equation}
Therefore, this model has no $\lambda_L = 0$ phase at small but nonzero $T,J$. 

To derive Eq.~\eqref{eq:main}, we improve upon the standard approach followed by \textit{ibid}, and analyze the leading eigenvalue problem of the ladder kernel, Eq.~(9) of \textit{ibid.}:
\begin{subequations} \label{eq:K}
\begin{align}
 &   K = K_2 K_1  \,,\,  (K_1 f)(t) = \left[ \kappa^2 + 3 J^2 G_{lr}^2 (t) \right]f(t) \,, \\
  &  (K_2 f)(\omega) = 
  \left|G_R\left(\omega + \im \frac{\lambda_L}{2}\right)\right|^2f(\omega) \,,
\end{align}
\end{subequations}
where $G_R$ and $G_{lr}$ are retarded and Wightman Green functions of a fermion, respectively. Recall also that $\lambda_L$ is found by imposing that the largest eigenvalue of $K$ is $1$. We shall work perturbatively in $J/\kappa$ and $T/\kappa$, up to the leading order. Such a perturbative expansion is controlled since the quartic term is irrelevant in the low-$T$ limit where the model is a Fermi liquid. Our analysis thus differs from \textit{ibid.}, which considered a perturbative expansion with respect to the relevant term.

To start, we compute $G_{lr}$ and $G_R$ to an adequate order. For $G_{lr}$, suffices the conformal solution of the $J = 0$ limit~\cite{syk}:
\begin{equation}
   G_{lr} (t)^2 = \frac{T^2}{\kappa^2}  \sech( t \pi T)^{2} \,. \label{eq:Glr}
\end{equation}
For $|G_R|^2$, the $J=0$ conformal solution $G_R(\omega) = -\im / \kappa$~\cite{syk} is not enough, and we shall incorporate the leading correction. That comes from quasiparticle decay, encoded in the imaginary part of the self-energy $\Sigma_{4}(\tau) := J^2 G(\tau)^3$ in Euclidean time. Continuing to real time, we find that the retarded propagator $\Sigma_{4}^R$ satisfies 
\begin{equation}
   -\mathrm{Im} \, \Sigma_{4}^R(\omega) =  \frac{J^2 T^2}{2\kappa^3} +\frac{J^2 \omega^2}{2\pi^2 \kappa^3} \,. \label{eq:sigma}
\end{equation}
at leading order in $J/\kappa$. Together with the Schwinger-Dyson equation $G_R(\omega)^{-1} = \omega - \kappa^2 G_R(\omega) - \Sigma_{4,R}(\omega)$, Eq.~\eqref{eq:sigma} implies that,
\begin{equation}
   \kappa^2  \left|G_R\left(\omega + \im \frac{\lambda_L}{2}\right)\right|^2 = 1-\frac{\lambda_L}{2\kappa}-\frac{J^2 T^2}{2\kappa^4} - \frac{J^2}{2\pi^2 \kappa^4}\omega^2 \,, \label{eq:GR2}
\end{equation}
where we also omitted  higher orders of $\lambda_L$, anticipating it to be small. 

With Eqs.~\eqref{eq:Glr} and \eqref{eq:GR2}, the kernel \eqref{eq:K} reduces to 
\begin{align}
    K = &1- \frac{\lambda_L}{2\kappa} - \frac{J^2 T^2}{2\kappa^4} -  \nonumber \\
    & \frac{J^2T^2}{\kappa^4}\left(- \frac{1}{2\pi^2 T^2} \partial_t^2 - 3  \sech( t \pi T)^{2}  \right)\,.
\end{align}
Upon rescaling $s = t\pi T$ in the parenthesis, we recognize the 1D Schr\"odinger Hamiltonian with a P\"oschl-Teller potential: $-\frac12 \partial_{s}^2 - 3 \sech(s)^2$. Its ground state energy is $E_0 = -2$~\cite{teller}. Thus the largest eigenvalue of $K$ equals $ 1-  {\lambda_L}/(2\kappa) + {3J^2 T^2}/(2\kappa^4). $
Imposing it to be $1$ leads to Eq.~\eqref{eq:main}. 

\begin{figure}
    \centering
    \includegraphics[width=0.75\columnwidth]{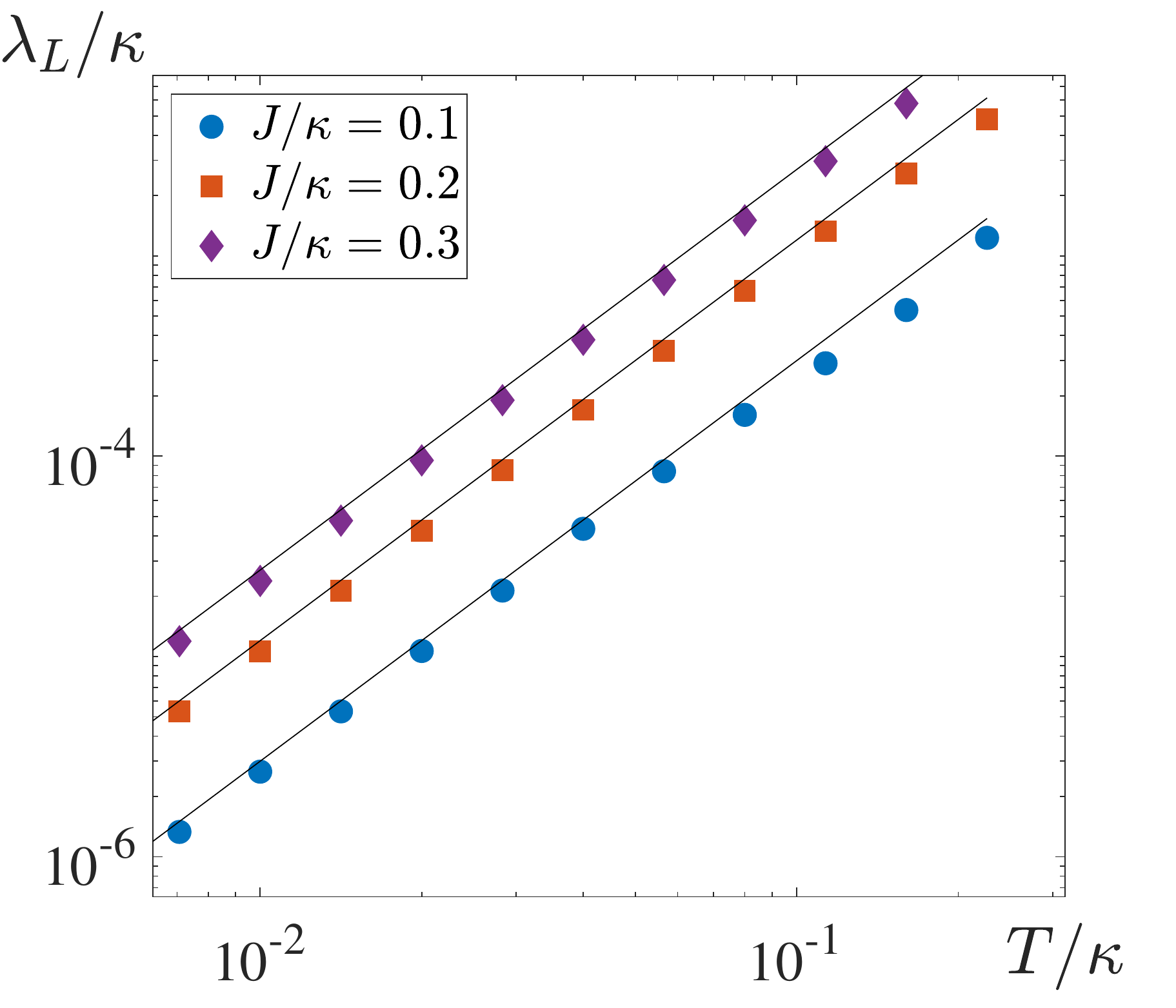}
    \caption{Temperature dependence of the Lyapunov exponent $\lambda_L$ for various interaction strengths (markers), compared to the prediction~\eqref{eq:main} (solid lines).}
    \label{fig:Num}
\end{figure}
We verified our prediction numerically in the large-$N$ limit, by directly computing the leading eigenvalue of the kernel~\eqref{eq:K}, where the Green functions are (non-perturbative) numerical solutions of the real-time Schwinger-Dyson equations~\cite{syk}. We find a nice agreement in the perturbative regime, see Fig.~\ref{fig:Num}. Observe however that $\lambda_L / \kappa$ can become very small, and could be mistaken for $0$ due to numerical artifacts. Finally, we note that Refs~\cite{sumilan,guo,sorokhaibam2019phase} observed $\lambda_L \sim T^2 > 0$ in other large-$N$ models of disordered Fermi liquid.  

 \begin{acknowledgements}
 	 We thank Jiabao Yang and Ehud Altman for helpful discussions. We acknowledge support from the DOE
 	 grant DE-SC0019380, and Gordon and Betty Moore Foundation's EPIC initiative, Grant GBMF4545 (X.C.).  \end{acknowledgements}

 %In conclusion, 

%with a P\"oschl–Teller potential. =
%We note that a $T^2$-dependence of the Lyapunov exponent has been observed numerically in similar lYet, to our knowledge, 

%Co
\bibliography{ref}
\end{document}